\documentclass[prb,preprint,showpacs]{revtex4}
\usepackage{graphicx}

\begin{document}
\title{Surface Spin-Flop Transition
in a Uniaxial Antiferromagnetic Fe/Cr Superlattice
 Induced by a Magnetic Field of Arbitrary Direction}
\author{M. G. Pini$^{1}$}
\email{mariagloria.pini@isc.cnr.it}
\author{A. Rettori$^{2,3}$}
\author{P. Betti$^2$}
\author{J. S. Jiang$^4$}
\author{Y. Ji$^4$}
\author{S. G. E. te Velthuis$^4$}
\author{G. P. Felcher$^4$}
\author{S. D. Bader$^4$}
\affiliation{
$^1$ Istituto dei Sistemi Complessi, CNR,
Sezione di Firenze,
Via Madonna del Piano, I-50019 Sesto Fiorentino (FI), Italy
\\
$^2$ Dipartimento di Fisica, Universit\`a di Firenze,
Via G. Sansone 1, I-50019 Sesto Fiorentino (FI), Italy
\\
$^3$ INFM-CNR National Research Center S3, Via Campi 213/A,
I-41100 Modena, Italy
\\
$^4$ Materials Science Division, Argonne National Laboratory, Argonne, IL 60439, USA
}
\date{\today}
\begin{abstract}
We studied the transition between the antiferromagnetic and the
surface spin-flop phases of a uniaxial antiferromagnetic [Fe(14
\AA)/Cr(11 \AA]$_{\rm x20}$ superlattice.  For external fields
applied parallel to the in-plane easy axis, the layer-by-layer
configuration, calculated in the framework of a mean-field
one-dimensional model, was benchmarked against published polarized
neutron reflectivity data.
For an in-plane field $H$ applied at
an angle $\psi \ne 0$ with the easy axis, magnetometry shows that
the magnetization $M$ vanishes at $H=0$, then increases slowly
with increasing $H$. At a critical value of $H$, a finite jump in
$M(H)$ is observed for $\psi<5^{\rm o}$, while a smooth increase
of $M$ $vs$ $H$ is found for $\psi>5^{\rm o}$. A dramatic increase
in the full width at half maximum of the magnetic susceptibility
is observed for $\psi \ge 5^{\rm o}$.  The phase diagram obtained
from micromagnetic calculations displays a first-order transition
to a surface spin-flop phase for low $\psi$ values, while the
transition becomes continuous for $\psi$ greater than a critical
angle, $\psi_{\rm max} \approx 4.75^{\rm o}$. This is in fair
agreement with the experimentally observed results.
\end{abstract}
\pacs{75.70.-i, 75.50.Ee, 75.10.-b, 75.30.Kz}

\maketitle

\section{Introduction}

It is well known that when a magnetic field applied along the easy
axis of a uniaxial antiferromagnet exceeds a critical value
$H_{BSF}=\sqrt{2H_E H_A+H_A^2}$, where $H_E$ is the exchange field
and $H_A$ is the anisotropy field,  the system undergoes a
first-order phase transition to a bulk spin-flop phase,
characterized by sublattice magnetizations nearly perpendicular to
the field direction.\cite{neel,callen} In the case of a uniaxial
antiferromagnet with one or two surfaces, which break the
translational invariance in the direction perpendicular to the
surface plane, the problem of determining the ground-state spin
configuration in the presence of an external magnetic field
applied along the in-plane easy axis was theoretically posed a few
decades ago. The first model to be investigated was that of a
semi-infinite stack of ferromagnetic planes, antiferromagnetically
coupled and subject to a magnetic field antiparallel to the
magnetization of the surface plane. For this system, when the
ratio $r$ between $H_A$ and $H_E$  is very small
($r=H_A/H_E\ll1$), the onset of a surface spin-flop phase was
predicted using a continuum approximation.\cite{mills,keffer} This
phase is characterized by a canting near the surface and is stable
for field values $H$ greater than a critical value
$H_{SSF}=\sqrt{H_E H_A+H_A^2}\approx H_{BSF}/\sqrt{2}$. Nearly a
decade ago, the existence of such a surface spin-flop phase was
criticized\cite{map,review} because a discrete, nonlinear map
approach showed that the instability of the antiferromagnetic
configuration at $H_{SSF}$ simply leads to an interchange of the
two sublattices.\cite{map,review} Subsequently Pokrovsky and
Sinitsyn\cite{pokrovsky} showed that, for a semi-infinite film
with $r\ll1$, a quite similar result can be obtained also in the
continuum approximation, provided that appropriate boundary
conditions are assumed.

The case of a finite stack with an even number $N$ of planes
(but still $r\ll1$) is quite different since, for $H_{SSF}<H<H_{BSF}$,
the system tends to realize an inhomogeneous configuration with
the magnetizations of both surfaces parallel to the field direction,
and a domain wall is thus located in the center of the stack.
For $H>H_{BSF}$ the stable state of the finite film is a different
inhomogeneous configuration, with the inner spins assuming nearly
the bulk spin-flop configuration and the surface ones less deviated
with respect to the field direction.\cite{review}

Fullerton {\it et al.}\cite{baderprb}  showed that the Fe/Cr(211)
superlattice, obtained by the repetition of $N$ ferromagnetic iron
layers antiferromagnetically coupled through the Cr spacer,
constitutes an excellent model system of a finite uniaxial
antiferromagnetic film. Since then, a great number of
papers\cite{wangprl,wmprb,mapchaos,micheletti,trallori,dantas,tevelthuis,rossler1,rossler2,rossler3,rbarxiv}
have been devoted to the study of the surface spin-flop phase
transition. In fact, for sufficiently low thickness of the Fe
layers, the system possesses a dominant uniaxial in-plane
anisotropy along the Fe[ 0 -1 1] direction, with $H_A$ of the same
order of magnitude as $H_E$. For example, the Fe/Cr(211)
superlattice\cite{wangprl} with thickness $t_{\rm Fe}=40$ \AA~ and
$t_{\rm Cr}=11$ \AA~ had an anisotropy-to exchange ratio $r= H_A/
H_E\approx 1/4$, while for the system investigated in the present
work and in Ref. \onlinecite{tevelthuis}, with $t_{\rm Fe}=14$
\AA~and $t_{\rm Cr}=11$ \AA, one has $r \approx 1/10$. This is a
major difference with respect to ordinary antiferromagnets, like
MnF$_2$, where $r$ is usually much smaller ($r \approx 1/100$).
The consequence of an increased value of $r$ in superlattices with
respect to bulk antiferromagnets was investigated both
experimentally\cite{wangprl} and
theoretically\cite{wmprb,mapchaos} in Fe/Cr(211) superlattices
with $r \approx 1/4$, and was found to introduce a sequence of
sudden jumps in the field dependence of the magnetization, in
addition to the surface spin-flop jump occurring at $H_{SSF}$.

Recently, an accurate and systematic study of the phase diagram of a
uniaxial antiferromagnetic film with an even number of planes was
performed by R\"o\ss ler and Bogdanov\cite{rbarxiv} using an
efficient conjugate gradient minimization technique in the case
of the external magnetic field applied precisely along the easy axis.
For $r\ll 1$, they found that there is only a first-order transition
from the collinear antiferromagnetic (AF) phase to a symmetric,
inhomogeneously flopped phase  with the spin-flop (SF) located
in the center of the film. For $r \le 1$, their calculations of
the spin configuration confirmed previous theoretical
findings.\cite{wangprl,mapchaos,micheletti} In fact they found
a series of canted, asymmetric phases (C$_{\rm i}$), separated
by first-order transition lines, between the AF and the SF
phases.\cite{rbarxiv} Within these intermediate C phases, the
ground state of the system evolves from a canted configuration
with a flop localized near one of the surfaces (C$_{\rm 1}$) to
other configurations (C$_{\rm 2}$, C$_{\rm 3}$, ...) where the
flop moves into the center, causing abrupt variations of the
magnetization as the field intensity is increased.\cite{rbarxiv,mapchaos}
Finally, upon further increasing the anisotropy ($r>1$), they
found that only first-order transitions between collinear
(antiferro-, ferri- and ferromagnetic) states are possible.\cite{rbarxiv}

While the magnetic phase diagram of the finite AF film has been
extensively investigated both theoretically and experimentally in
the case of $H$ applied parallel to the easy axis, no such studies
are known for the case for which $H$ is applied in-plane along an
arbitrary direction  with the easy axis. In the bulk case, the
field-induced phase transition of a uniaxial antiferromagnet in
the presence of a skew field forming an angle $\psi \ne 0$ with
the easy axis was theoretically studied by Rohrer and
Thomas\cite{rohrer} using a mean-field approach and then by Fisher
{\it et al.}\cite{fisher,kosterlitz} Neglecting zero-point motion
effects, they determined the equilibrium configurations as a
function of  the skew field, and found that the phase boundary
between the AF and the bulk SF phase extends only to a maximum
angle with respect to the easy axis, where it ends in critical
points. More precisely, they predicted the first-order SF
transition to become continuous ({\it i.e.}, second-order) for $\psi$
greater than a critical angle $\psi_{\rm max}$(bulk)$=\tan^{-1}
[H_A/(2H_E-H_A)]$. This expression, first developed for a
tetragonal system, was later found to be valid for an orthorhombic
system, when $\psi$ is restricted to a plane comprising the easy
and intermediate axes.\cite{blazey}

The small values of $\psi_{\rm max}$(bulk) in ordinary bulk
antiferromagnets (amounting to a few tenths of a degree)
made the observation of the crossover from first- to
second-order in the transition difficult. Early results
on MnF$_2$ were, in this respect, only qualitative.
More direct evidence of the crossover character of the
transition was provided by Butera {\it et al.}\cite{butera}
by measuring the staggered magnetization of of MnCl$_2\cdot$4H$_2$O.
However, the existence and nature of a bicritical point was proven
by measuring the critical magnetic scattering of a number
of systems, notably CuCl$_2\cdot$2H$_2$O,\cite{lynn} CsMnBr$_3\cdot$2H$_2$O,\cite{bastee}
and the above mentioned MnCl$_2\cdot$4H$_2$O. On intuitive grounds,
one might expect that in Fe/Cr(211) superlattices, a similar
crossover effect in the order of the surface SF transition
should be present and should be more easily observable owing
the higher value of the ratio $r=H_A/H_E$  ($r \approx 1/10$ or more)
compared to ordinary bulk antiferromagnets like MnF$_2$, where $r \approx 1/100$.

The aim of the present paper is twofold: {\it i)} to extend to the
film ({\it i.e.} finite) case the theoretical study of the
magnetic phase diagram of a uniaxial antiferromagnet in a skew
field, and {\it ii)} to provide experimental evidence that, in
contrast with bulk antiferromagnets, in the case of the [Fe(14
\AA/Cr(11 \AA)]$_{\rm x20}$ superlattice, the crossover in the
order of the surface SF transition with increasing skew field
might be observed. Concerning the theoretical analysis, in Sec. II
we present two different approaches, the nonlinear map method and
the Landau-Lifshitz-Gilbert micromagnetic simulation, for the
ground state and full hysteresis calculations, respectively. In
Sec. III experimental results are presented and discussed,
obtained by different techniques, both for $H$ parallel to the
easy axis ($\psi=0$) and for a skew field ($\psi \ne 0$). Finally,
conclusions are drawn in Sec. IV.

\section{Film model and theoretical framework}

We consider a film made of an even number, $N$, of parallel
ferromagnetic planes that are antiferromagnetically coupled one to
each other by a nearest neighbor exchange interaction and subject
to a uniaxial in-plane anisotropy. For thin magnetic layers, the
magnetostatic dipolar interaction is known to confine the spins to
the film plane, so that, at equilibrium, the spins are necessarily
in-plane and the dipolar energy is zero. One can therefore
characterize the
spin configuration of the $i$-th plane by only one parameter,
namely the angle
$\phi_i$  that the magnetization of the $i$-th plane forms with
the $z$ axis ({\it i.e.}, with the easy anisotropy axis) within
the film plane $xz$.
A magnetic field $H$ is applied in the film
plane along a direction that forms an angle $\psi$ with $z$. At
$T=0$ K the energy density of the system takes the form:
\begin{eqnarray}
\label{energy} e&=& E/(g\mu_B SN_{\Vert}) \cr &=& \sum_{i=1}^{N}[
H_E \cos (\phi_i - \phi_{i-1}) -H_A \cos^2 \phi_i  \cr &-&2H_Z \cos
\phi_i -2 H_X \sin \phi_i ] ,
\end{eqnarray}
where $N_{\Vert}$ denotes the number of spins within each film layer,
and $H_Z=H \cos \psi$, and $H_X=H \sin \psi$.

The equilibrium spin configurations can be obtained from (\ref{energy}) by $\phi_i$-derivation ($i=1,\cdots,N$)
\begin{eqnarray}
\label{equilibrium}
{{\partial e}\over {\partial \phi_i}}=0=
&-& H_E \sin (\phi_i - \phi_{i-1}) (1- \delta_{1,i})
\cr &-& H_E \sin (\phi_{i+1} - \phi_i) (1- \delta_{N,i})
\cr
&+& 2H_A \sin \phi_i \cos \phi_i
\cr
&+& 2H_Z \sin \phi_i - 2 H_X \cos \phi_i
\end{eqnarray}

In the present work, the ground state of the one-dimensional (1D)
model (\ref{energy}) that approximates the film has been
determined using two different theoretical methods, namely: {\it
i)} an integration of the Landau-Lifshitz equation for the spin
chain, introducing a damping coefficient in order to reach the
magnetic ground state (see Sec. II.A), and {\it ii)} a
reformulation of Eq.~(\ref{equilibrium}), which provide the
equilibrium conditions in terms of a nonlinear map with opportune
boundary conditions at the film surfaces (see Sec. II.B). A
comparative discussion of the two methods is made in Sec. II.C.

\subsection{Spin configuration via integration of the Landau-Lifshitz-Gilbert equation}

The Landau-Lifshitz-Gilbert (LLG) equation\cite{landau}
\begin{eqnarray}
{{d{\bf M}}\over {dt}}= &-&\gamma({\bf M}\times {\bf H}_{\rm eff} )
\cr
&+&(\alpha_{\rm G} / M_s ) ( {\bf M} \times  {{d{\bf M}}\over {dt}})
\end{eqnarray}
describes the physical path of the magnetic moment
${\bf M}$ in a field ${\bf H}_{\rm eff}$. Here $\gamma$  and
$\alpha_{\rm G}$ are the gyromagnetic ratio of the free electron
spin and the Gilbert damping coefficient respectively, $M_s$ is
the saturation magnetization, and ${\bf H}_{\rm eff}  = -
(1/M_s)(\partial e/\partial {\bf M})$ is the local effective
magnetic field. For thin magnetic layers, a large value of
$\alpha_{\rm G}$  is appropriate because the demagnetizing field
confines the magnetic moment to the film plane, suppressing the
gyromagnetic precession and leaving an in-plane rotation of the
moment towards the direction of the local effective field. Thus,
the integration of the LLG equation becomes a simple relaxation of
the magnetic moment along the energy gradient. In the numerical
calculations, we iteratively rotate spins that represent
individual Fe layers by an amount proportional to the torque ${\bf
M} \times ({\bf M} \times{\bf H}_{\rm eff})$ at each instant. ${\bf
H}_{\rm eff}$  is then evaluated from the resulting configuration
and applied to the next iteration. Upon reaching convergence, the
stability of the equilibrium is tested by evaluating the
eigenvalues of the stability matrix $M_{ij}=\partial^2 e/(\partial
\phi_i \partial \phi_j)$.\cite{wmprb,dantas} All eigenvalues of
the stability matrix must be positive for the state to be stable.
In the event of an unstable equilibrium, the configuration
$\left\lbrace \phi_i \right\rbrace$ is displaced by a random, small fraction along the
eigenvector direction for which the eigenvalue is negative, and
the relaxation process starts anew until the system reaches a
stable {\it local} energy minimum.

\subsection{Determination of the energy minima via the nonlinear map method}

A different approach for the determination of the ground state of the
magnetic system described by
Eq.~(\ref{energy}) was proposed some years ago.\cite{map,mapchaos,review}
It is based upon a reformulation of the equilibrium conditions (\ref{equilibrium})
of the magnetic film model in terms of a discrete nonlinear map, where
the film surfaces are introduced via opportune boundary conditions.\cite{pandit}

We start by introducing the new variable $s_i= \sin (\phi_i - \phi_{i-1})$,
so that the conditions  for an equilibrium spin configuration,
Eq. (\ref{equilibrium}), can be rewritten as a 2D nonlinear recursive map. For $1<i<N$ one has
\begin{eqnarray}
\label{mappa}
s_{i+1} &=& s_i - (H_A/H_E) \sin (2\phi_i)
\cr &-& 2(H_Z/H_E) \sin \phi_i
\cr
&+& 2(H_X/H_E) \cos \phi_i ,
\cr
\cr
\phi_{i+1}&=& \phi_i + \sin^{-1}(s_{i+1}).
\end{eqnarray}
The trajectories in ($\phi, s$) space are associated with
equilibrium configurations, while the fixed points of the map
correspond to uniform ground states of the infinite system and are
second-order [{\it i.e.}, they satisfy the relation $(\phi_{n+2},
s_{n+2}) =(\phi_n, s_n)$] owing to its AF nature. We denote by
($\alpha, \beta$) the ground state configuration of the infinite
system in the presence of a field of arbitrary direction. The
angles $\alpha$ and $\beta$ that the magnetizations of the two
sublattices form with the easy axis can be determined by
numerically solving the following problem of extremum in 2D:
\begin{eqnarray}
\label{bulk}
\partial e/\partial \alpha =&0&=H_E \sin(\beta-\alpha)
+(H_A /2) \sin (2\alpha)
\cr &+& H_Z \sin \alpha -H_X \cos \alpha,
\cr
\cr
\partial e/\partial \beta =&0&=H_E \sin(\alpha-\beta)
+(H_A /2) \sin (2\beta)
\cr &+& H_Z \sin \beta -H_X \cos \beta.
\end{eqnarray}
From Eq.~(\ref{bulk}) it is readily found that the second-order
fixed points of the map are FP$_1$=($\beta,\sin (\beta-\alpha)$)
and FP$_2$=($\alpha,-\sin (\beta-\alpha)$). In order to study the
map behavior in the proximity of the fixed points, it is useful
to perform a linear stability analysis of the doubly iterated map.
It is worth noticing that energetically stable configurations
({\it i.e.}, with a positive definite Hessian) are associated
with topologically unstable ({\it i.e.}, hyperbolic) trajectories
in phase space. And, {\it vice versa}, energetically unstable
configurations are associated with topologically stable
({\it i.e.}, elliptic) trajectories in phase space.\cite{pandit}

At this point, the presence of the two surface planes, signaled
by the terms with the Kronecker's
$\delta$'s in Eq.~(\ref{equilibrium}), is taken into account
via opportune boundary conditions: {\it i.e.}, we introduce
two fictitious planes $i=0$ and $i=N+1$, so that we can assume
the bulk equations, Eq.~(\ref{mappa}), to be valid even for the
surface planes $i=1$ and $i=N$, provided that the following
equations are satisfied:
\begin{eqnarray}
\label{boundary}
s_1&=&\sin(\phi_1 -\phi_0)=0
\cr \cr
s_N&=&\sin(\phi_{N+1} -\phi_N)=0.
\end{eqnarray}
In this way, among all trajectories obtained from the map
equations (\ref{mappa}), only those satisfying the boundary
conditions (\ref{boundary}) represent equilibrium
configurations of the film with a finite number $N$
of planes. In practice, the physical trajectories of the
film must have two intersections with the $s=0$ line,
separated by exactly $N$ steps of the recursive mapping.\cite{map,mapchaos,review}

Using this nonlinear map method, one is able to numerically
determine all the stationary configurations of the film
very rapidly and within machine precision. To find the
ground state among the various calculated equilibrium
configurations, it is necessary first to perform a linear
stability analysis of the obtained solutions through the
evaluation of the Hessian and then to compare the energies
of the different metastable states in order to choose the
lowest energy one.

\subsection{Comparison of the two methods}

By definition, the two theoretical methods described in Sec.~II.A and II.B
give the same results for the field evolution of the ground state from the
AF to the SF phase. For the sake of generality, a comparison between the
two methods and a critical discussion of their respective advantages and
drawbacks will now be presented.

The nonlinear map method allows one to determine all the
equilibrium configurations of the film at zero temperature. Thus,
comparing the energies of the various equilibrium configurations,
one can easily determine the ground state. However, in the
presence of two or more energetically equivalent ground states of
different configurations, their knowledge is not enough to predict
when a given equilibrium state is abandoned in favor of another
equilibrium state as the intensity of the external field is
varied. Such a process cannot be simulated in the framework of the
nonlinear map method, whereas, in the approach based on the
integration of the Landau-Lifshitz-Gilbert equation, the inclusion
of the damping coefficient allows the system to evolve towards a
stable local energy minimum. It should be noted that, in the
framework of the latter method, the
magnitude of perturbation applied during the stability test ({\it
i.e.} the fractional displacement of the configuration
$\left\lbrace \phi_i \right\rbrace$ in the event of an unstable
equilibrium) may be crucial in determining which one of these
states is eventually reached.

To this regard, we note that the problem of the determination of
the actual state of the system may be nontrivial in the case of
uniaxial anisotropy field and exchange field with comparable
intensities, $r=H_A/H_E \approx 1$, as in Fe/Cr(211) superlattices
for a suitable choice of the layer thicknesses.\cite{wangprl}
In fact the metastability region turns out to be strongly
amplified with respect to the case $r\ll 1$, and many metastable
states with slightly different energies may be present, so that
a tiny increment in the value of $H$ may cause an abrupt change
in the ground-state configuration.\cite{mapchaos,rossler1,rossler2}
Such a peculiar feature is usually signaled by a chaotic aspect
of the nonlinear map in the ($\phi, s$) phase space.\cite{mapchaos,review}
In the extreme case of $r\gg 1$, fractal structures were predicted
to appear both in the distribution of magnetic moments and
in the energy spectrum.\cite{kurten}

As regards the [Fe(14 \AA)/Cr(11 \AA)]$_{\rm x20}$ superlattice
under study, in Section III it will be shown  that this system is
characterized by a moderate value of the anisotropy ($r\approx
1/10$). As a consequence, the metastable state to which the system
eventually relaxes in the micromagnetic calculations is not very
sensitive to the magnitude of the perturbation applied during the
stability test, so long as it is reasonably small. Thus, owing to
the low value of $r$, the two theoretical methods were found to
give quite similar results for the ground state of the Fe/Cr
superlattice. (For the sake of precision, the results reported in
Fig.~1 were obtained by the approach based upon the integration of
the LLG equation, while those in Fig.~2, Fig.~6 and Fig.~7 were
obtained via the nonlinear map method.)

Finally, it is worth observing that the theoretical results refer
to the ground state, while experiments (as described in Section
III) are performed at room temperature $T_{amb}$. The influence of
a finite temperature on the spin-flop transition of a classical,
simple cubic lattice, uniaxially anisotropic Heisenberg
antiferromagnet was investigated some time ago.\cite{fisher}
Roughly speaking, comparing the $T=0$ results with the
experimental ones at $T_{amb}$ is reasonable as long as $T_{amb}
\lesssim T_b$, the bicritical point where the AF and SF ordered
phases meet in the $(H,T)$ phase diagram. For the square lattice,
recent work\cite{holt} indicated that a very narrow disordered
phase may intervene between the AF and the SF phase down to quite
low temperatures, leading to the definition of a tetracritical
point $T_t>T_b$. In both the $D=3$ and $D=2$ cases, the
multicritical point in the $(H,T)$ phase diagram was found to be a
substantial fraction of $T_N$, the AF-paramagnetic transition
temperature for $H=0$. The Fe/Cr superlattice under study cannot
be described by either of the two afore-mentioned models, since it
is made of strongly ferromagnetic iron films,
antiferromagnetically coupled through the chromium spacer.
However, denoting by $T_{mc}$ the multicritical point in the
$(H,T)$ phase diagram, one can expect $T_{mc} \lesssim T_N$ also
in the superlattice case. In fact, on the basis of a mean field
theory estimate, $T_N$ is expected to be much greater than
$T_{amb}$ for the Fe/Cr superlattice, so that one can guess also
the condition $T_{amb} \lesssim T_{mc}$ to be satisfied.

\section{Experimental results}

\subsection{Sample characterization}

The preparation and characterization of epitaxial Fe/Cr
superlattices are similar to those described  in Ref.~\onlinecite{baderprb}.
The [Fe(14 \AA)/Cr(11 \AA)]$_{\rm x20}$
superlattice was prepared\cite{tevelthuis} by {\it dc} magnetron
sputtering onto a single-crystal MgO(110) substrate. To assure
epitaxy with the substrate, a 200-\AA~buffer layer of Cr was first
deposited at 400 $^{\rm o}$C, then the superlattice was deposited
at 100 $^{\rm o}$C and found to grow with a (211) orientation.
Finally, a 100-Å\AA~capping layer of Cr was deposited to protect
the sample. The epitaxy and the smoothness of the superlattice
were checked by x-ray diffraction and found to have an interfacial
roughness of $\approx 4$ \AA. Extensive magnetic characterizations were
performed by means of magnetometry, as well as by
magnetoresistance measurements. For Fe film thickness $t_{\rm
Fe}=14$ \AA, a strong, in-plane surface anisotropy $K_S=0.06$
ergs/cm$^2$ was found to develop along the [0 -1 1] direction,
leading to a uniaxial in-plane anisotropy $K_U=2K_S/t_{\rm
Fe}=8.6~10^5$ ergs/cm$^3$ (compared to the bulk crystalline
anisotropy $K_1=4.7~ 10^5$ ergs/cm$^3$). Using the value
$M_s=1740$ emu/cm$^3$ of bulk Fe at $T=0$ K, one obtains
$H_A=2K_U/M_s=985.2$ Oe. The coupling between ferromagnetic layers
was found to oscillate as a function of the thickness of the Cr
interlayer. For $t_{\rm Cr}=11$ \AA~an AF exchange coupling with
strength $J_{\rm AF}= -1.194$ ergs/cm$^2$ was
estimated,\cite{baderprb} leading to $H_E=2\vert J_{\rm
AF}\vert/(t_{\rm Fe}M_s)=9802.9$ Oe.

For the [Fe(14 \AA)/Cr(11 \AA)]$_{\rm x20}$ superlattice   under
study, previous SQUID measurements\cite{tevelthuis} of $M(H)$ $vs$
$H$, applied in-plane along the easy axis of the sample, showed
that $M$ is zero for zero field. As $H$ was increased, the
instability of the AF phase was signaled by a steep increase of
$M(H)$ at $H_{SSF}\approx 2.73$ kOe. In this study, $M$ was
measured using a vibrating sample magnetometer (VSM), in which the
sample can be rotated so as that the in-plane magnetic field is
applied at any skew angle  $\psi$ with respect to the easy axis.

\subsection{Spin configuration for H parallel to the easy axis ($\psi=0$)}

For $H$ applied in-plane parallel to the easy axis ($\psi=0$), the
calculation of the ground-state magnetization profiles shows that,
in a limited field range near $H_{SSF}=\sqrt{H_E H_A+H_A^2}$, the
system admits two stationary configurations: the AF state, with
collinear and antiparallel layer magnetizations, and the surface
spin-flop (SSF) state, with a non-uniform magnetization profile
characterized by a Bloch wall that nucleates near one of the film
surfaces. In Fig.~1 we plot, in the neighborhood of $H_{SSF}$, the
field dependence of the reduced magnetization, $m(H)=M/M_s=(1/N)
\sum_{i=1}^N \cos \phi_i (H)$, and of the reduced energy $e(H)$,
given by Eq.~(\ref{energy}). Using for the calculations $H_E=9.80$
kOe, $H_A=0.98$ kOe, $N=20$ and $\psi=0$,  the field of
thermodynamic equivalence between the AF and the SSF state is
found to be $H_{\rm th}=H(e_{\rm AF}=e_{\rm SSF})=3.02$ kOe, while
the boundaries of the metastability region are given by the fields
$H_{\rm inf}=2.93$ kOe and $H_{\rm sup}=H_{SSF}=3.26$ kOe. The AF
state is the ground state for $H<H_{\rm th}$ and is metastable for
$H_{\rm th}<H<H_{\rm sup}$, while the SSF state is the ground
state for $H>H_{\rm th}$ and is metastable for $H_{\rm
inf}<H<H_{\rm th}$. The calculated value of the bulk SF field is
$H_{BSF}=\sqrt{2H_E H_A+H_A^2} =4.50$ kOe.

In Figs.~2,a-d the calculated ground-state magnetization profiles
are compared with those obtained from polarized neutron
reflectivity measurements, as published in
Ref.~\onlinecite{tevelthuis}, at different field values, ranging
between 0 and 5.5 kOe. The fields $h=H/H_{BSF}$ were scaled with
respect to the bulk SF field, where we set $H_{BSF}=4.50$ kOe for
the theoretical results and $H_{BSF} \approx 4.14$ kOe for the
experimental ones. The orientation $\phi_{AF}(i)$ of the
antiferromagnetic axis ({\it i.e.}, the axis along which the
magnetizations of two Fe adjacent layers are antiparallel) is
plotted as a function of the Fe layer number for selected values
of the applied field. Figure 2e illustrates that
$\phi_{AF}(i)={1\over 2} [\phi(i)+\phi(i-1)]-90^o$, where
$\phi(i)$ is the angle formed by the magnetization ${\bf M}_i$ of
the $i$-th Fe layer with the field direction. As the measured
spectrum extends only as far as the half order AF Bragg peak,
which is determined by the antiparallel components of the
magnetizations, the orientation $\phi_{AF}(i)$ of the AF axis is
obtained with more accuracy from the experiments than that of the
individual layer orientations, for which the estimated error is up
to 20$^o$.\cite{tevelthuis} This representation clearly depicts
the position and extent of the domain wall for the different
fields and shows that agreement between theory and experiment is
fairly good.

From Figs.~2,a-d one sees that the basic features of the SSF
transition in this film, characterized by a low value of $r
\approx 1/10$, are the following: {\it i)} for $H>H_{SSF}$ the
deviations from the uniform AF spin configuration originate just
at the surface layer whose magnetization is antiparallel to the
field; {\it ii)} with increasing $H$, the surface-nucleated domain
wall is pushed gradually into the middle of the film; {\it iii)}
for $H>H_{BSF}$ a symmetric spin configuration is achieved,
similar to the bulk SF one in the middle planes (while the spins
at the surfaces, owing to the cuts of the exchange bonds, are less
deviated from the field direction with respect to the bulk ones).
Although not directly obvious from Fig.~2, the
discommensuration\cite{micheletti} at the center of the
surface-nucleated domain wall, effectively dividing the AF order
into two antiphase domains, which was strongly evidenced by the
neutron data,\cite{tevelthuis} is also reproduced by the
calculations.

Note that no abrupt variations of the magnetization, except the
one at $H_{SSF}$, are found as the field intensity is increased,
see Fig.~1: {\it i.e.}, additional first-order transitions between
C phases, intermediate between the AF and the SSF one,\cite{rbarxiv,mapchaos}
are not allowed by the low value of $r=H_A/H_E$ in this system.

\subsection{Spin configuration for H applied along an arbitrary direction ($\psi \ne 0$)}

For a field applied in-plane along an arbitrary direction forming
a skew angle $\psi \ne 0$ with the easy axis ({\it i.e.}, both
$H_X$ and $H_Z$ are non zero), one has that $M(H)=0$ for $H=0$.
For $0<\psi<5^{\rm o}$,  $M(H)$ increases slowly with increasing
$H$, since for $\psi \ne 0$ the magnetizations of the two
sublattices are no longer compensated. Upon further increasing
$H$, a finite jump, signaling the onset of a first-order phase
transition, was observed for a $\psi$-dependent field value. The
measurements were performed both upon increasing and decreasing
$H$ and showed a marked hysteresis for such small $\psi$ values:
see Fig.~3. The magnetic susceptibility $\chi(H)=dM/dH$, obtained
by numerically deriving the measured magnetization with respect to
$H$, showed sharp peaks corresponding to the jumps in $M(H)$: see
Fig.~4, top. The full width at half maximum (FWHM, see later
Fig.~5, top) of the measured susceptibility peak was found  to be
constant and very small (essentially determined by the
instrumental resolution), signaling that for such small angle
values ($0<\psi<5^{\rm o}$) the SSF transition is of first order.

As $\psi$ was gradually increased above 5$^{\rm o}$, $M(H)$ became
smoother (see Fig.~3), and the peak in $\chi(H)$ (see Fig.~4,
bottom) decreased in intensity, while the FWHM dramatically
increased on passing from $5^{\rm o}$ to $23^{\rm o}$, as shown in
Fig.~5, top. The latter feature strongly suggests that a crossover
of the surface phase transition from first-order to second-order
might take place for $\psi \ge 5^{\rm o}$.

In the light of this interpretation, it is however necessary to
justify the persistence - up to the highest investigated value of
$\psi$ (23$^{\rm o}$) - of a small hysteresis loop: see Fig.~5,
bottom, where the measured dependence of the peak position of the
magnetic susceptibility $\chi(H)$ is shown as a function of the
skew angle $\psi$, both for increasing (full circles) and
decreasing (open circles) magnetic field.

To this aim, we observe that the coexistence of first- and
second-order transition features was recently observed\cite{li} in
single-crystal La$_{0.73}$Ca$_{0.27}$MnO$_3$ perovskites
exhibiting colossal magnetoresistance. The magnetization isotherms
displayed a metamagnetic structure linked with a first-order
transition, while field and temperature dependent $ac$
susceptibility data presented a crossover line characteristic of a
continuous transition.\cite{li}

In our case of a finite AF Fe/Cr(211) film, a similar effect, {\it
i.e.} the coexistence of first- and second- order transition
features, might be attributed to a distribution of values of the
interlayer exchange, as well as of the anisotropy of the different
layers in the stack, due to the presence of thickness
fluctuations. In determining the observed small hysteresis loop,
one cannot either rule out the role of defects (pinning centers)
which inhibit the lateral motion of domains during the
magnetization reversal process.

In the following we will test if the experimental data can be
explained in terms of a crossover from first- to second-order
critical behavior by performing a theoretical calculation of the
magnetization profile $\left\lbrace \phi_i; i=1, \cdots, N
\right\rbrace$ for different values of $H$ and $\psi \ne 0$, using
either of the two methods described in Sec. II. In the present
case where $r\ll 1$, the map portrait is not chaotic and the two
different methods give similar results. From the calculated spin
configuration, one obtains $m(H)=M/M_s=(1/N)\sum_{i=1}^N \cos
\phi_i$, $\chi(H)=dM/dH$ and $e(H)$, given by Eq. (\ref{energy}).

We find that for $0<\psi<4.75^{\rm o}$ the system admits
two stationary configurations: a nearly AF state and a SSF
state. The first one is stable only for $H<H_{\rm sup}$
while the second one is stable only for $H>H_{\rm inf}$.
By $H_{\rm th}$ we denote the field of thermodynamic
equivalence at which the two states take the same energy.
As $\psi$ increases, the width of the metastability region
gradually reduces until, for $\psi>4.75^{\rm o}$, only one
equilibrium configuration is found.

In Fig. 6 the calculated susceptibility  $\chi(H)$ is shown for
different values of the skew angle $\psi$. For $\psi \le 3^{\rm
o}$, the peak in $\chi(H)$ is a Dirac delta function, so the peak
position is indicated by a vertical line. For $\psi \ge 5^{\rm o}$
the peak has a finite width and a finite height. As $\psi$
increases, the peak broadens and its height decreases. For
clarity's sake, the peak position reported in Fig.~6 for $\psi \le
3^{\rm o}$ corresponds to the calculated field of thermodynamic
equivalence $H_{\rm th}$ between the energies of the AF-like
configuration and the non-uniform SSF configuration. It is just
the position of this peak that is reported {\it vs} $\psi$ in
Fig.~7 as the full-circle diagram; the other two diagrams plotted
for $\psi \le 4.75^{\rm o}$ represent the calculated field values
$H_{\rm inf}$ (open squares) and $H_{\rm sup}$ (open triangles)
{\it vs} $\psi$.

In the phase diagram of Fig.~7 one clearly observes that, upon
increasing $\psi$, the width of the metastability region gradually
shrinks until, for $\psi$ greater than a critical value
$\psi_{\rm max}$(film)=4.75$^{\rm o}$, the film admits
only one equilibrium state. Thus, for $\psi \ge \psi_{\rm max}$(film)
we expect the SSF transition to become continuous. The calculated
value of $\psi_{\rm max}$(film)=4.75$^{\rm o}$ turns out to be
in remarkable agreement with the value $\approx 5^{\rm o}$
estimated from the experimental results on the basis of the
strong increase observed in the FWHM of $\chi(H)$. This fact
provides support for the hypothesis of a crossover from first-
to second-order critical behavior for the SSF transition in a skew field.

The calculated value of $\psi_{\rm max}$(film)=4.75$^{\rm o}$
should also be compared with its bulk counterpart. In the bulk
case, the field-induced phase transition of a uniaxial
antiferromagnet in the presence of a skew field forming
an angle $\psi$ with the easy axis was theoretically
studied by Rohrer and Thomas.\cite{rohrer} They predicted the
first-order bulk SF transition to become continuous
for $\psi \ge \psi_{\rm max}$(bulk)=$\tan^{-1} [H_A/(2H_E-H_A)]$.
However, in  MnF$_2$, where $r \approx 1/100$,  the critical angle
turns out to be as small as $\approx 0.4^{\rm o}$. In the case of
the Fe/Cr superlattice  under study, the ratio $r$ is nearly an
order of magnitude higher than in MnF$_2$, so that an appreciable
critical angle $\psi_{\rm max}$(bulk)$\approx 3^{\rm o}$ is estimated.
The calculated value of $\psi_{\rm max}$(film)=4.75$^{\rm o}$ is
nearly twice the bulk value. This can be qualitatively understood
considering that, for not too high values of $r$, in the bulk the
critical angle is essentially determined by the ratio $H_A/(2H_E)$,
while in the film the effective exchange field at the surface is
halved with respect to the bulk.

\section{Discussion}

In this work we have investigated the transition, induced by a
magnetic field $H$ with arbitrary direction, between the
antiferromagnetic phase and the surface spin-flop phase of an
epitaxial Fe/Cr(211) superlattice with $t_{\rm Fe}=14$ \AA, $t_{\rm Cr}=11$ \AA~and
$N=20$ repetitions. The system is characterized by a rather
small value ($r \approx 1/10$) of the ratio $r=H_A/H_E$
between the uniaxial anisotropy field $H_A$ and the exchange
field $H_E$, yet much greater than the value ($r \approx 1/100$)
pertinent to usual bulk antiferromagnets like MnF$_2$ and Cr$_2$O$_3$.

For an external field applied parallel to the easy in-plane axis
($\psi=0$), the layer-by-layer spin configurations measured by
polarized neutron reflectometry were found to be in remarkable
agreement with theoretical calculations, performed in the
framework of a mean-field 1D model of the superlattice stack.

For a field applied in-plane along an arbitrary direction forming
a skew angle $\psi \ne 0$ with the easy axis, the superlattice
magnetization $M(H)$ was measured using magnetometry. The phase
diagram of the film was calculated in order to check the possibility
of a crossover of the surface phase transition from first- to
second-order to take place for $\psi_{\rm max}$(film), similarly
to what was predicted decades ago by Rohrer and Thomas\cite{rohrer}
for an AF bulk system in a skew field. Indeed we
calculated $\psi_{\rm max}$(film)$\approx 4.75^{\rm o}$, to be
compared with the experimental value, $\psi \approx 5^{\rm o}$,
at which the jump in $M(H)$ starts smoothening and the FWHM of
the measured magnetic susceptibility displays a dramatic increase
with increasing $\psi$. Owing to the cut of exchange bonds at
the film surfaces, the calculated value of $\psi_{\rm max}$(film)
turns out to be nearly twice its bulk counterpart,
$\psi_{\rm max}$(bulk)$\approx 3^{\rm o}$.\cite{rohrer}
The latter value is much higher than the ones predicted for
ordinary bulk antiferromagnets ({\it e.g.},
$\psi_{\rm max}$(bulk)$\approx 0.4^{\rm o}$ for MnF$_2$ and
$\approx 0.015^{\rm o}$ for Cr$_2$O$_3$).\cite{rohrer}

From the comparison between our experimental and theoretical
results we conclude that a crossover between first- and
second-order critical behavior is easier to be observed
by magnetization measurements in Fe/Cr superlattices, thanks
to the much higher value of the ratio $r=H_A/H_E$ between
the anisotropy and the exchange fields in such an artificially
grown system with respect to ordinary bulk antiferromagnets.

The interpretation of the experimental data proposed above needs,
however, further investigation to be conclusive. In particular, a
quantification of the magnetic domain structure in the presence of
a structurally rough interface is required. While our previously
published\cite{tevelthuis} polarized neutron reflectometry data
proved unambiguously that only one type of domains is present in
Fe/Cr(211) superlattices with an interfacial roughness of $\approx
4~$\AA~for zero field, such a clear-cut evidence is lacking in the
case of a nonzero field, applied in plane along an arbitrary
direction. In conclusion, while we do not claim a quantitative
accuracy for our theoretical results, nevertheless we believe that
the main features of the spin-flop transition in the Fe/Cr
superlattice have been captured by our "ideal" model ({\it i.e.},
characterized by structurally smooth and uniformly magnetized
layers).

\begin{acknowledgments}
Financial support from the Italian Ministery for University and
Research is acknowledged. This work was performed in the framework
of the joint CNR-MIUR programme (legge 16/10/2000, Fondo FISR) and
of the COFIN Project on Magnetic Multilayers. The work at Argonne
National Laboratory was supported by the US Department of Energy,
Office of Science, under contract W-31-109-ENG-38.
\end{acknowledgments}

\newpage
\begin{figure}
\includegraphics[width=10cm,angle=0,bbllx=118pt,bblly=253pt,%
bburx=490pt,bbury=731pt,clip=true]{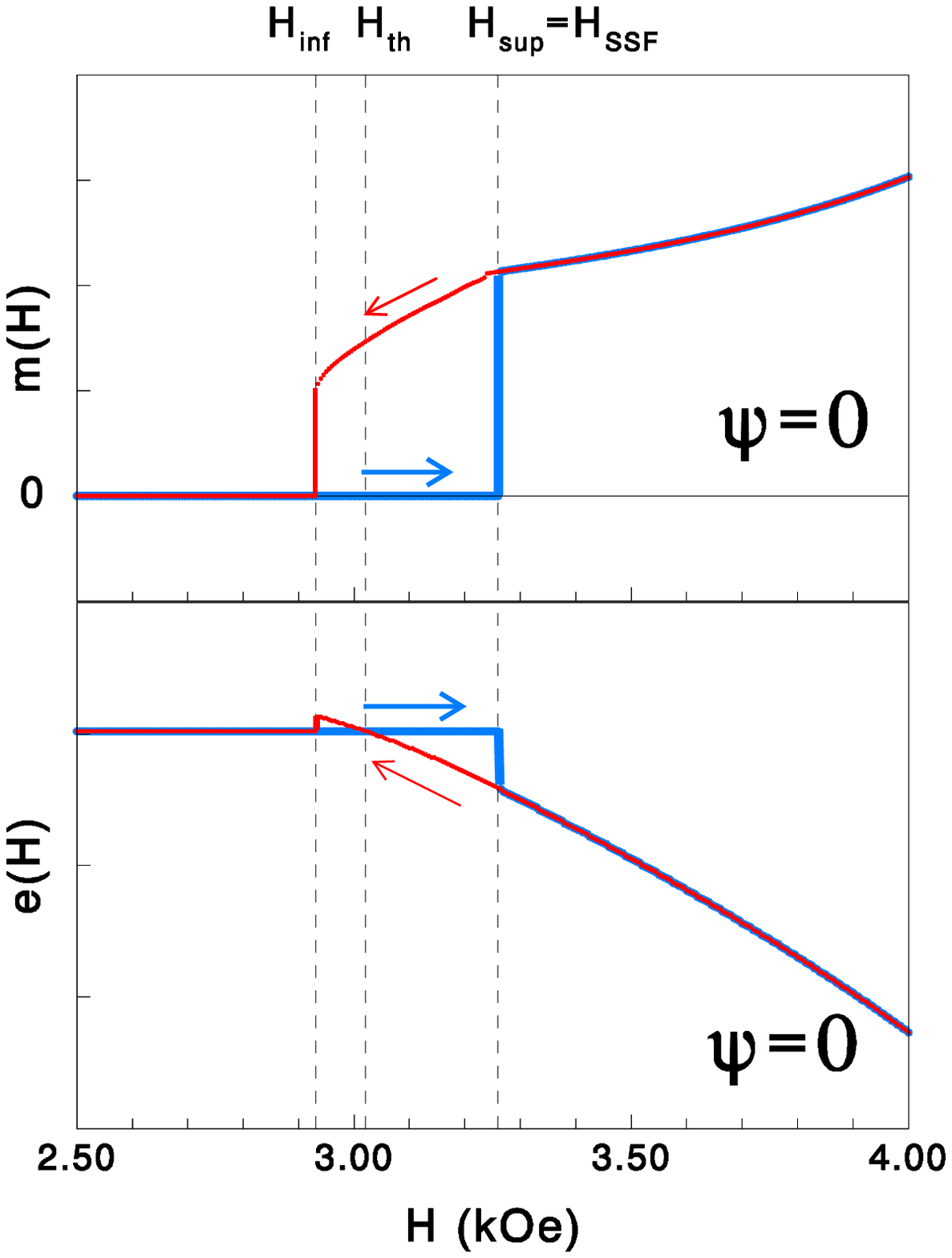} \caption{(Color online)
Calculated field dependence of the reduced magnetization,
$m(H)=M/M_s$, and reduced energy $e(H)$, given by Eq.~(1), of an
antiferromagnetic film with $N=20$ planes, for $H$ in the
neighborhood of the surface spin-flop transition. The field is
applied along the easy axis ($\psi=0$); the exchange field and the
anisotropy field are, respectively, $H_E=9.80$ kOe and $H_A=0.98$
kOe. The thick (thin) line refers to increasing (decreasing)
magnetic field. The field of thermodynamic equivalence between the
energies of the collinear AF and the non-uniform SSF configuration
is $H_{\rm th}=3.02$ kOe. The collinear AF state is stable for
$H<H_{\rm sup}=3.26$ kOe; the non-uniform SSF state is stable for
$H>H_{\rm inf}=2.93$ kOe; they have the same energy at $H_{\rm
th}$. }
\end{figure}

\begin{figure}
\includegraphics[width=10cm,angle=0,bbllx=75pt,bblly=264pt,%
bburx=511pt,bbury=700pt,clip=true]{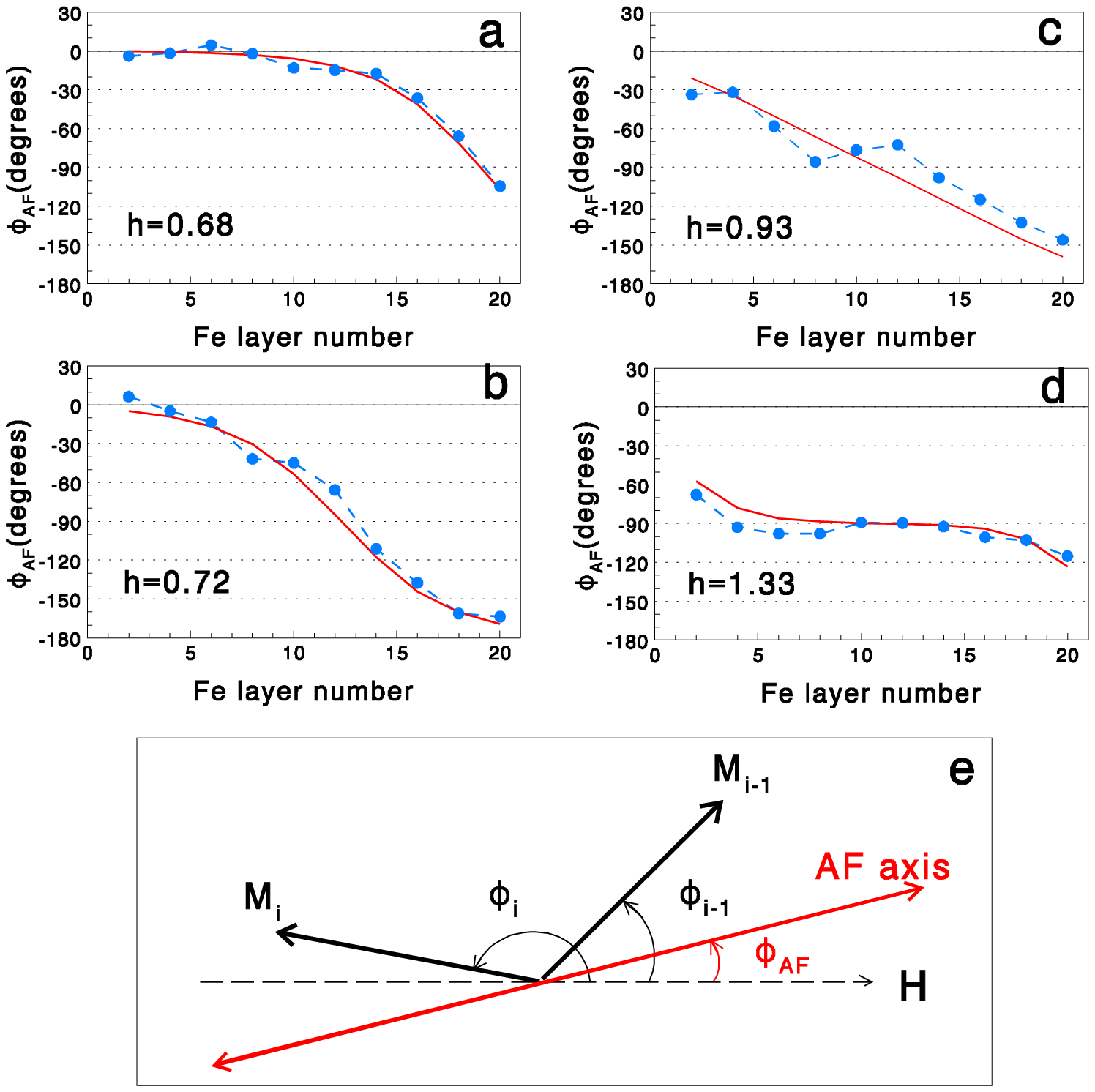} \caption{(Color online)
a-d) Comparison between the calculated layer-by layer
configuration (solid lines) and the experimental ones (markers and
dashed lines) deduced from polarized neutron reflectometry data in
a [Fe(14 \AA)/Cr(11 \AA]$_{\rm x20}$ superlattice.
The orientation $\phi_{AF}$ of the antiferromagnetic (AF) axis
({\it i.e.}, the axis along which the magnetizations of two
adjacent Fe layers are antiparallel) is plotted as a function of
the Fe layer number for selected values of the reduced applied
field ($h=H/H_{BSF}$); e) Schematic drawing showing that
$\phi_{AF}(i)={1\over 2} [\phi(i)+\phi(i-1)]-90^o$, where
$\phi(i)$ is the angle formed by the magnetization ${\bf M}_i$ of
the $i$-th Fe layer with the field direction. The magnetic field
is applied parallel to the easy axis ($\psi=0$).}
\end{figure}

\begin{figure}
\includegraphics[width=10cm,angle=0,bbllx=93pt,bblly=259pt,%
bburx=513pt,bbury=552pt,clip=true]{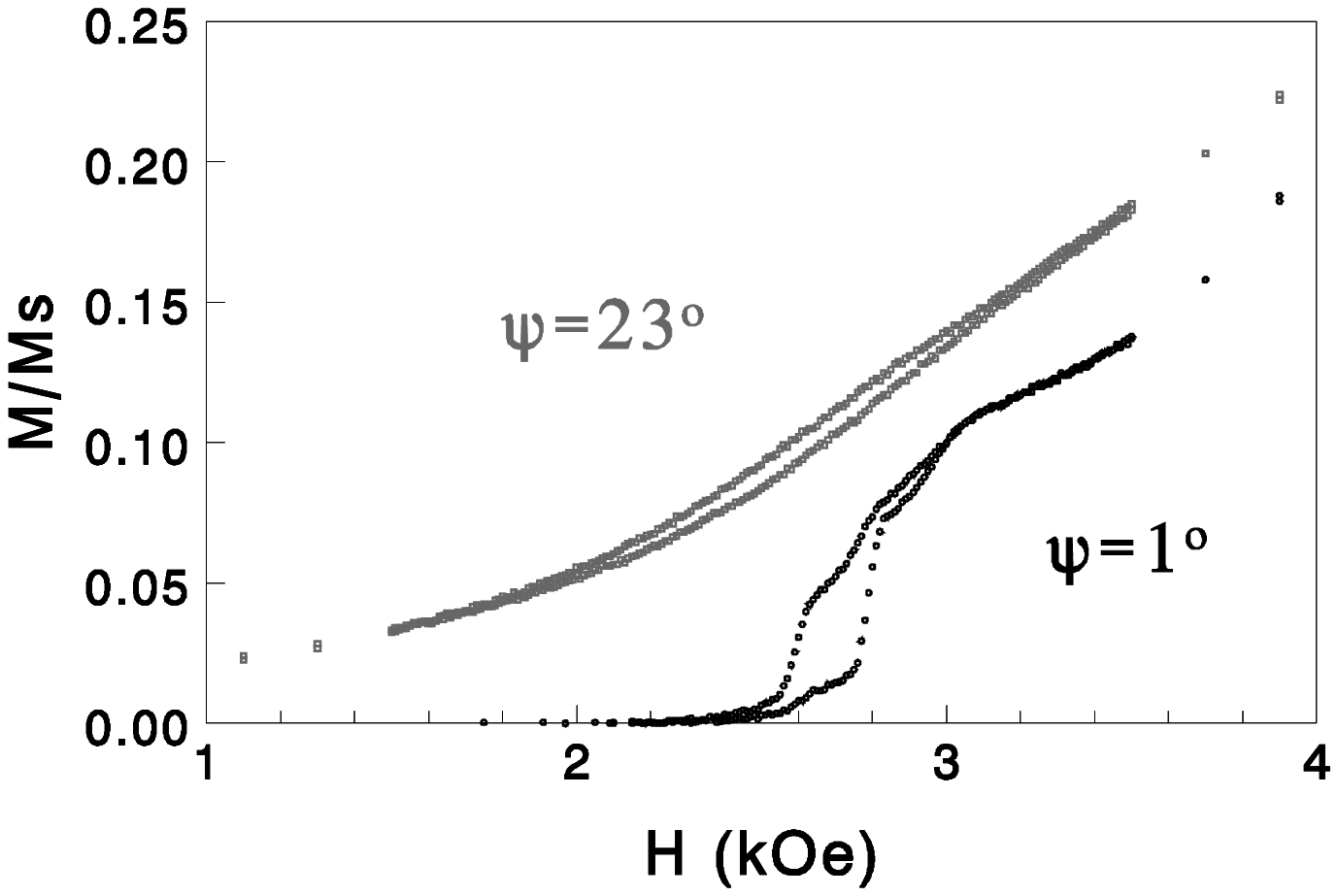} \caption{Experimental
VSM data for the magnetization $M(H)$ of a [Fe(14 \AA)/Cr(11
\AA)]$_{\rm x20}$ superlattice $vs$ $H$, applied along a direction
which forms a skew angle $\psi$ with the easy in-plane axis. The
two different curves refer to $\psi=1^{\rm o}$ and $\psi=23^{\rm
o}$, respectively.}
\end{figure}

\begin{figure}
\includegraphics[width=10cm,angle=0,bbllx=105pt,bblly=235pt,%
bburx=500pt,bbury=701pt,clip=true]{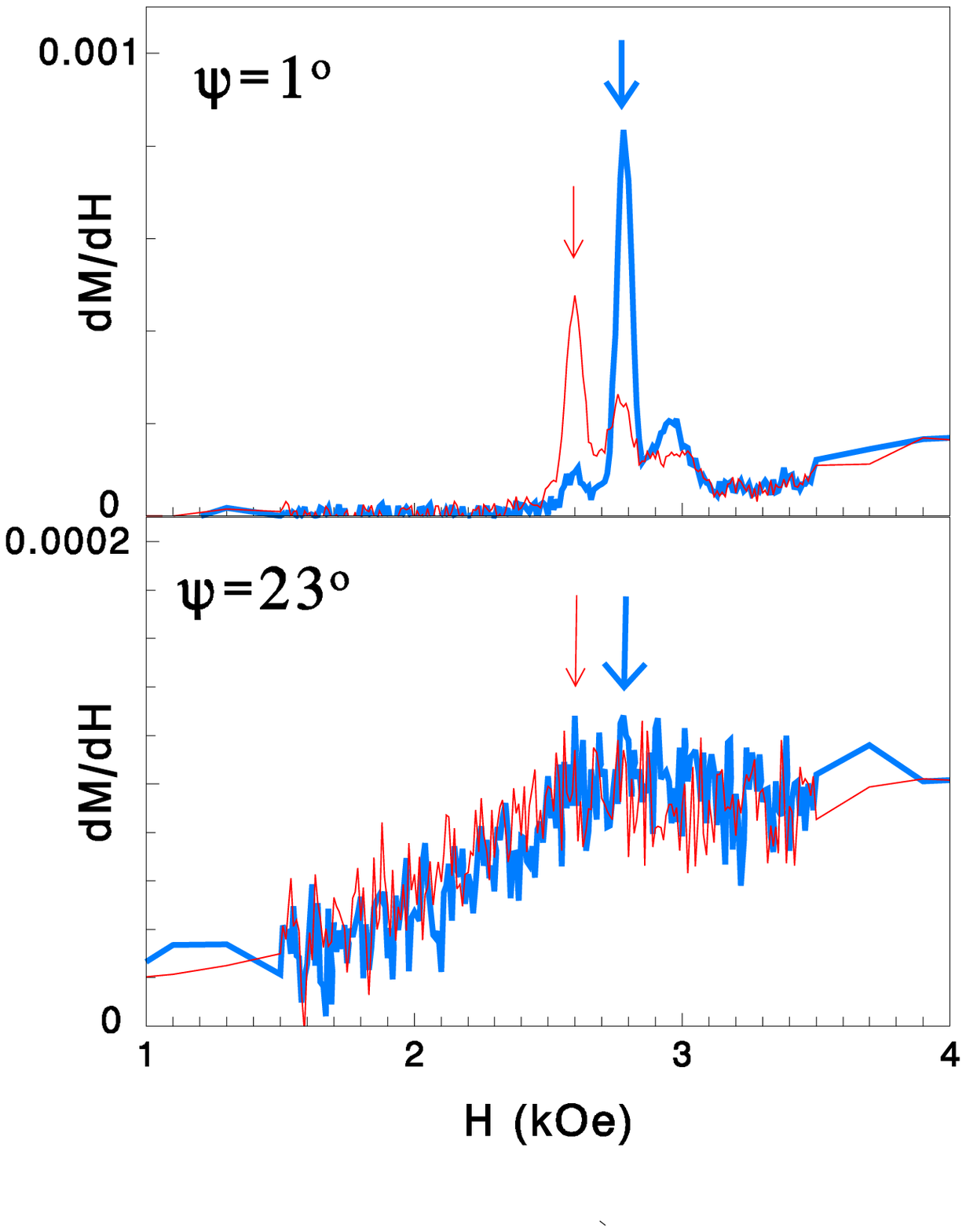} \caption{(Color online)
Experimental data for the magnetic susceptibility $\chi(H)=dM/dH$
of a [Fe(14 \AA)/Cr(11 \AA)]$_{\rm x20}$ superlattice, obtained by
numerically deriving the VSM data in Fig.~3, both for increasing
field  (thick line) and decreasing field (thin line). Top diagram:
$\psi=1^{\rm o}$ ; bottom diagram: $\psi=23^{\rm o}$. In each
diagram, the arrow at the lower (higher) field denotes the peak
position for decreasing (increasing) $H$.}
\end{figure}

\begin{figure}
\includegraphics[width=10cm,angle=0,bbllx=193pt,bblly=238pt,%
bburx=420pt,bbury=555pt,clip=true]{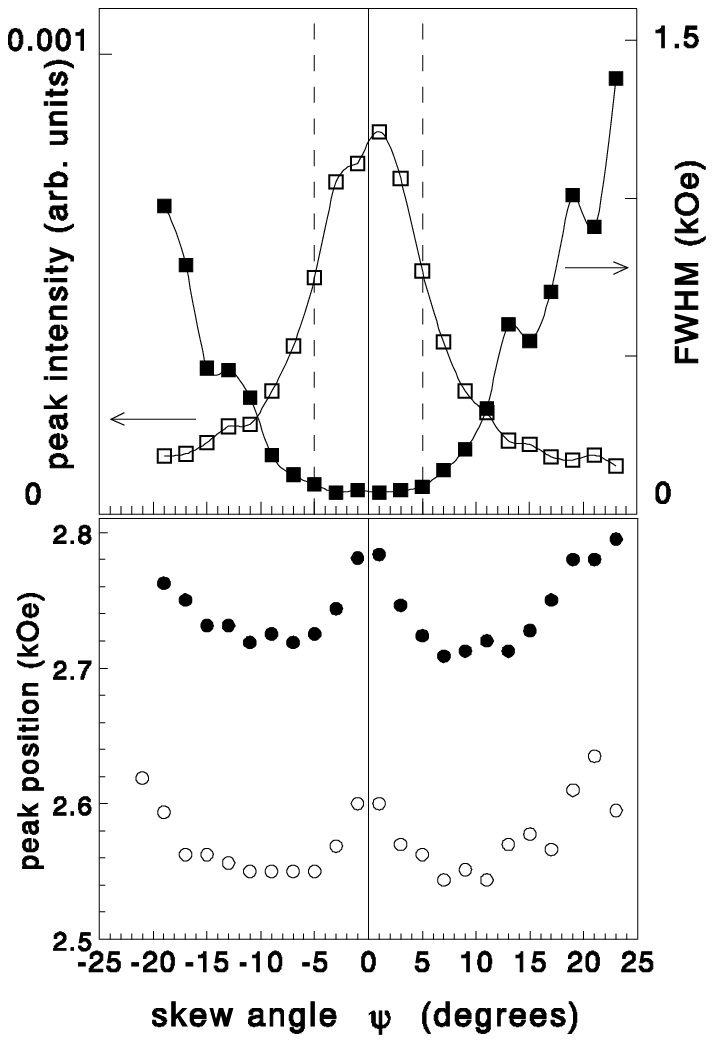} \caption{Top diagram:
experimental peak intensity (open squares) and FWHM (full squares)
of the magnetic susceptibility $\chi(H)$ of a [Fe(14 \AA)/Cr(11
\AA)]$_{\rm x20}$ superlattice $vs$ the skew angle $\psi$ formed
by the applied magnetic field with the easy in-plane axis.  The
lines are guides to the eye. A dramatic increase in the FWHM is
observed for $\vert \psi \vert > 5^{\rm o}$. Bottom diagram:
experimental peak position of $\chi(H)$ $vs$ $\psi$, both for
increasing (full circles) and decreasing (open circles) field
intensity.}
\end{figure}

\begin{figure}
\includegraphics[width=10cm,angle=0,bbllx=104pt,bblly=265pt,%
bburx=510pt,bbury=549pt,clip=true]{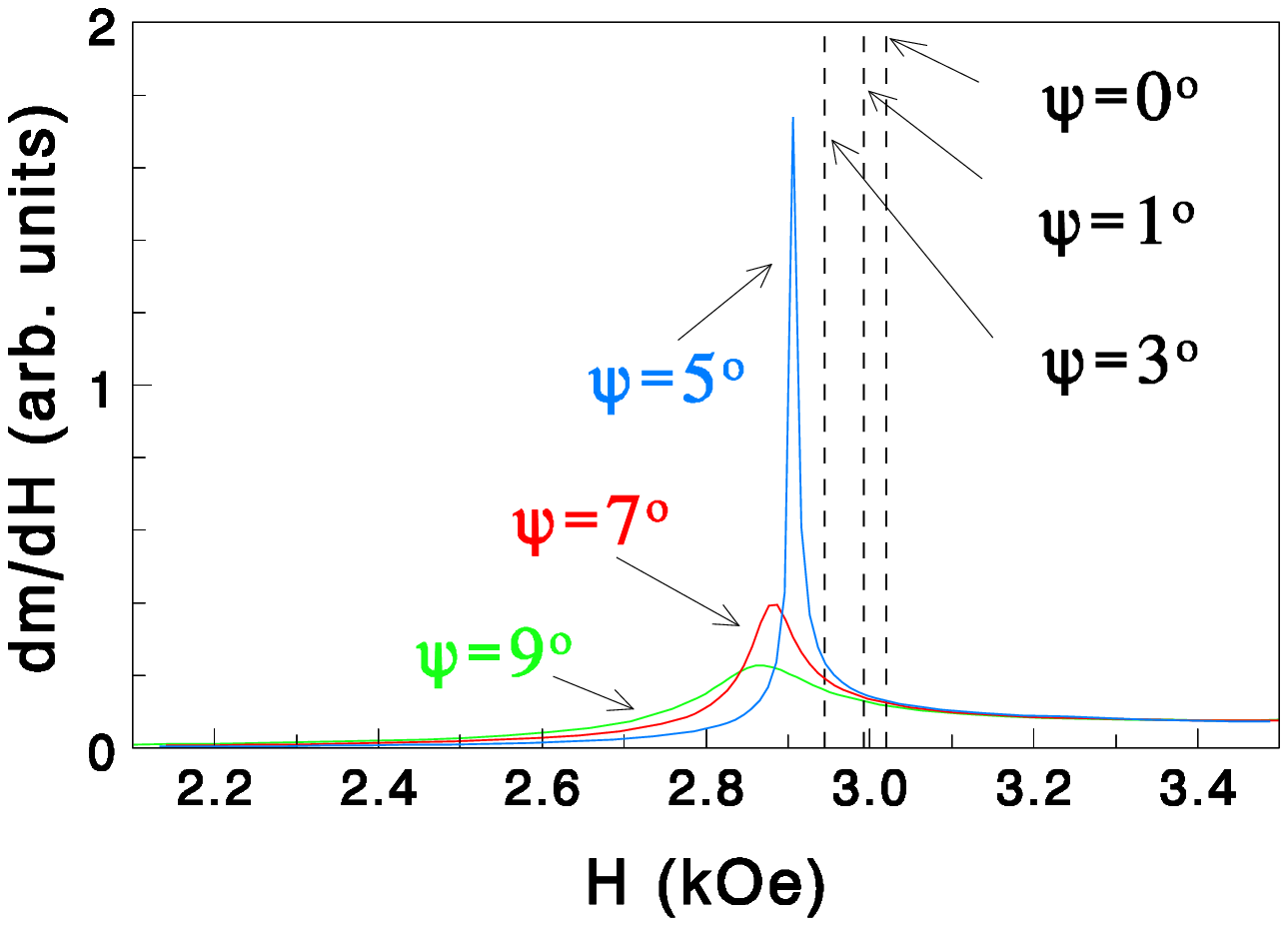} \caption{(Color online)
Calculated $\chi(H)=dM/dH$ of an AF film with $N=20$ and
$H_E=9.80$ kOe, $H_A=0.98$ kOe $vs$ $H$ applied in-plane along an
arbitrary direction. The different curves refer to different
values of the skew angle $\psi$ formed by the external magnetic
field with the easy axis. For $\psi \le 3^{\rm o}$, the position
of the reported peak of $\chi(H)$ is indicated by a vertical
dashed line and corresponds to the field of thermodynamic
equivalence $H_{\rm th}$ between the energies of the AF-like and
the non-uniform SSF configuration.}
\end{figure}

\begin{figure}
\includegraphics[width=10cm,angle=0,bbllx=81pt,bblly=212pt,%
bburx=547pt,bbury=544pt,clip=true]{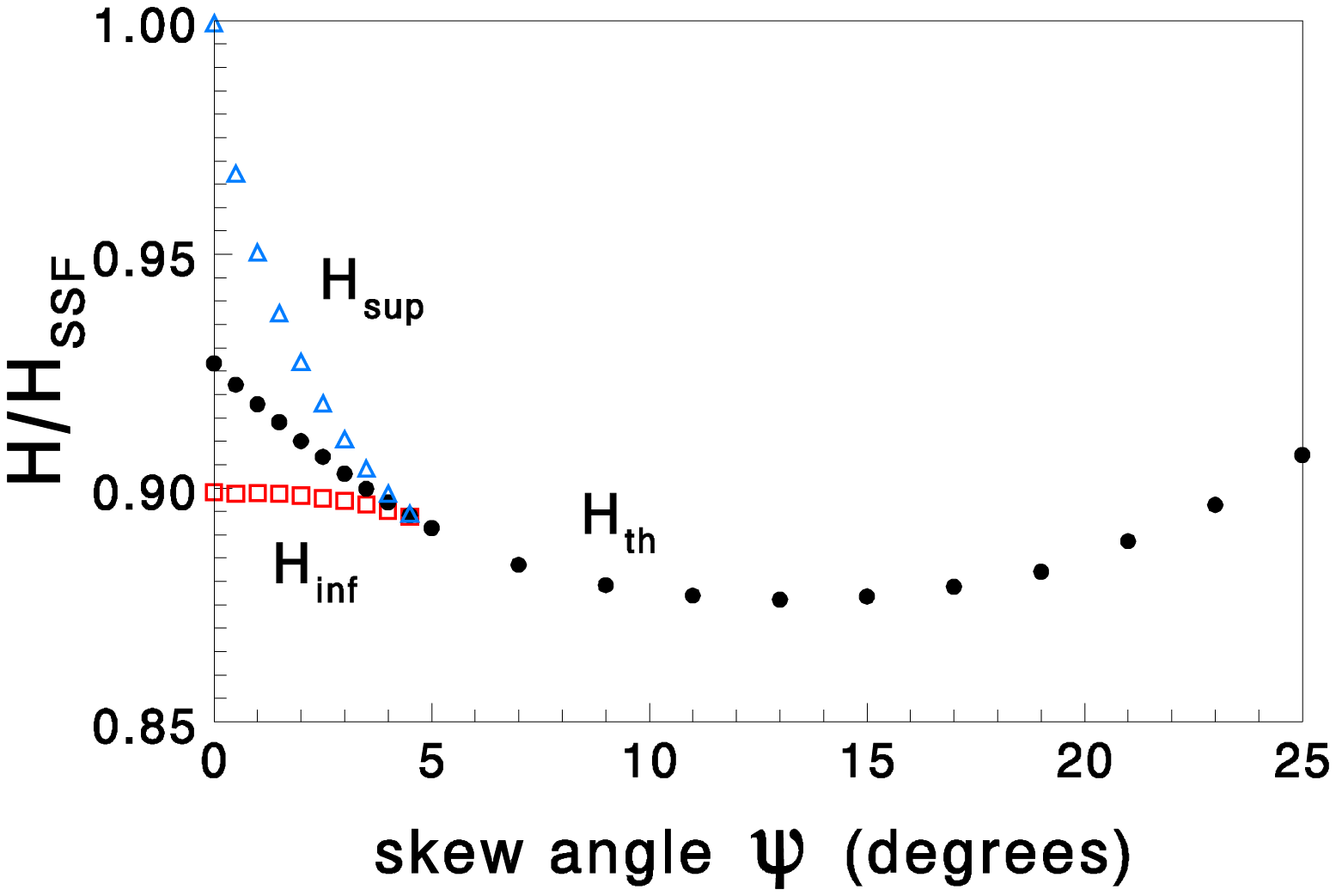} \caption{(Color online)
Calculated phase diagram of an AF film with $N=20$ and $H_E=9.80$
kOe, $H_A=0.98$ kOe. The peak positions of $\chi(H)=dM/dH$, scaled
with respect to $H_{SSF}$, are reported vs the skew angle $\psi$
formed by the external magnetic field with the easy axis. For
$\psi>4.75^{\rm o}$ the SSF transition is predicted to become
continuous. The calculated field of thermodynamic equivalence
$H_{\rm th}$ is reported {\it vs} $\psi$ as the full-circle
diagram; the other two diagrams plotted for $\psi \le 4.75^{\rm
o}$ represent the calculated field values $H_{\rm inf}$ (open
squares) and $H_{\rm sup}$ (open triangles).}
\end{figure}

\end{document}